\RequirePackage{fix-cm}
\documentclass[twocolumn,epjc3]{svjour3}  
\RequirePackage{graphicx}
\usepackage{graphics}
\usepackage{bm}
\usepackage{amsmath}
\usepackage{amsfonts}
\usepackage{amssymb}
\usepackage{comment}
\usepackage{physics,empheq,bm}
\usepackage{color}

\begin{document}
	\title{Mechanical and geometrical properties of jammed wet granular materials}
	\author{Kiwamu Yoshii\thanksref{e1,addr1} and Michio Otsuki\thanksref{addr1}}
	\thankstext{e1}{e-mail: k\_yoshii@fm.me.es.osaka-u.ac.jp}
	\institute{Department of Mechanical Science and Bioengineering, Graduate School of Engineering Science, Osaka University, 1--3 Machikaneyama, Toyonaka, Osaka 560--8531, Japan\label{addr1}}
	
	\date{Received: date / Accepted: date}
	
	\abstractdc{
		We numerically investigate the mechanical and geometrical properties of dense wet granular particles with irreversible attractive interaction.
		The shear modulus exhibits two inflection points as the packing fraction increases, and the bulk modulus shows a non-monotonic behavior.
		The coordination number also exhibits two inflection points.
		The peak position in the pair correlation function shifts to a lower value due to attractive interaction.
		The Voronoi tessellation of the particle configuration reveals that the probability density function for the volume of the Voronoi cell broadens as the packing fraction approaches the jamming point.
	}
	
	\maketitle
	\section{Introduction}\label{sec:intro}
	
	Amorphous materials, such as granular materials, emulsions, foams, and colloidal suspensions, behave like solids with rigidity when the packing fraction $\phi$ exceeds a critical volume fraction $\phi_c$, while they exhibit liquid-like behaviors for $\phi<\phi_c$.
	This transition, known as the jamming transition,
	has been extensively studied for years \cite{liu1998jamming,liu2010jamming,van2009jamming,behringer2018physics}.

	The mechanical properties of the materials exhibit critical behaviors near the critical fraction $\phi_c$.
	The pressure $P$, the shear modulus $G$, and the bulk modulus $B$ for repulsive frictionless particles exhibit power law scalings as a function of $\phi - \phi_c$ \cite{ohern2002random,ohern2003jamming}.
	Critical scaling laws for the rheological properties have been observed for systems under steady shear \cite{olsson2007critical,hatano2007criticality,hatano2008scaling,tighe2010model,otsuki2009critical}.
	Recently, the nonlinear elasticity of jammed amorphous materials \cite{coulais2014shear,otsuki2014avalanche,boschan2016beyond,otsuki2022softening}, the effect of friction \cite{somfai2007critical,silbert2010jamming,otsuki2017discontinuous,otsuki2021shear}, and the frequency dependence of the complex shear modulus \cite{tighe2011relaxations,dagois2017softening} have been studied.

	The geometrical properties also change drastically in the vicinity of $\phi_c$.
	When the packing fraction exceeds $\phi_c$, the coordination number $Z$ of frictionless particles changes from zero to the isostatic value $Z_{\mathrm{iso}}$.
	The excess coordination number $Z- Z_{\mathrm{iso}}$ exhibits a power law scaling as a function of $\phi - \phi_c$ \cite{ohern2002random,ohern2003jamming}.
	Moreover, in the three-dimensional systems consisting of frictionless mono-dispersed spheres with diameter $d$,
	the pair correlation function $g(r)$ diverges as $r \to d$ at $\phi_c$, which indicates that a lot of particles are on the verge of making contact \cite{silbert2006structural}.

	Earlier studies on the jamming transition have focused on purely repulsive particles.
	However, cohesion, such as the capillary force in wet granular materials, is nonnegligible in various realistic situations. 
	It is well known that the irreversible capillary force affects the dynamics of granular materials \cite{herminghaus2005dynamics,strauch2012wet,herminghaus2013wet,mitarai2006wet}, which might result from the change in their mechanical and geometrical properties.
	Some researchers have studied
	the rheological properties of cohesive particles
	in systems at constant volume \cite{chaudhuri2012inhomogeneous,gu2014rheology,irani2014impact,irani2016athermal,irani2019discontinuous} or 
	under constant pressure \cite{rognon2008dense,khamseh2015flow,yamaguchi2018rheology,badetti2018shear,mandal2021rheology,vo2020additive,vo2020evolution,vo2020role,macaulay2021viscosity}. 
	The yield stress or the apparent friction coefficient increases due to the attractive interaction between particles.
	The attractive force also causes a gel-like contact network for low $\phi$ \cite{head2007well,zheng2016shear}, shear bands \cite{irani2014impact,irani2016athermal,irani2019discontinuous,singh2014effect}, and clusters of particles \cite{yamaguchi2018rheology,vo2020evolution,macaulay2019shear,lois2008jamming}. 
	For two-dimensional particles with a simple reversible attractive force,
	it is reported that
	the transition point $\phi_c$ for rigidity decreases as the strength of cohesion increases \cite{koeze2018sticky},
	and the critical scaling laws in $G$ and $B$ for repulsive particles are broken \cite{Koeze2020Elasticity}.
	However, the behavior of three-dimensional wet granular materials with the irreversible capillary force near $\phi_c$ remains unclear.
	
	In this study, we numerically investigate the mechanical and geometrical properties of three-dimensional frictionless wet granular materials.
	In Sect. \ref{Setup}, we explain our model and setup.
	Section \ref{sec:mechanical} presents the numerical results on the mechanical properties.
	In Sect. \ref{sec:geometrical}, we deal with the geometrical properties.
	Section \ref{sec:geometrical} consists of three parts.
	The $\phi$-dependence of the coordination number is shown in Sect. \ref{sec:Z}.
	We demonstrate the change in the pair correlation function for $\Gamma_s>0$ in Sect. \ref{sec:Pair}.
	In Sect. \ref{sec:Voro}, we analyze the geometry of the particle configuration using the Voronoi tessellation.
	Section \ref{sec:Conclusion} presents discussion and conclusions of our results.
	The dependence of the critical fraction $\phi_c$ on the strength of cohesion is shown in \ref{sec:phic}.
	In \ref{sec:critical}, we investigate the effect of the capillary force on scaling laws for $G$.

	
	\section{Setup}\label{Setup}
	We consider three-dimensional wet granular materials consisting of $N$ frictionless identical particles in a cubic box of linear size $L$. 
	We apply oscillatory shear under Lees-Edwards boundary conditions using the SLLOD method to measure the mechanical properties \cite{evans2008non}.
	The equation of motion is given by
	\begin{align}
		\dfrac{\dd \bm{r}_i}{\dd t} &= \dfrac{\bm{p}_i}{m} + \Dot{\gamma}(t) y_i \bm{e}_x, \label{eq:eom1}\\
		\dfrac{\dd \bm{p}_i}{\dd t} &= \sum_{j \neq i} F_{ij} \bm{n}_{ij}- \Dot{\gamma}(t) p_{i,y} \bm{e}_x \label{eq:eom2}
	\end{align}
	with the mass $m$, the unit vector along the x-direction $\bm{e}_x$, and the shear rate $\Dot{\gamma}(t)$.
	Here, $\bm{r}_{i} = (x_i, y_i, z_i)$ and $\bm{p}_{i} = m_i(\frac{d}{dt}{\bm{r}}_i-\Dot{\gamma}(t)y_i)\bm{e}_x$ are the position and peculiar momentum of particle $i$, respectively.
	The force between particles $i$ and $j$ is denoted by $F_{ij}$.
	The force $F_{ij}$ consists of the dissipative force $F_{ij}^{(\mathrm{d})}$ and the static force $F_{ij}^{(\mathrm{s})}$ as
	\begin{align}
		F_{ij} = F_{ij}^{(\mathrm{d})} + F_{ij}^{(\mathrm{s})}.
	\end{align}
	The normal unit vector $\bm{n}_{ij}$ is given by $\bm{n}_{ij} = \bm{r}_{ij}/r_{ij}$ with $\bm{r}_{ij} = \bm{r}_{i} - \bm{r}_{j}$ and $r_{ij} =  |\bm{r}_{ij}|$.

	The dissipative force is given by
	\begin{align}
		F_{ij}^{(\mathrm{d})} &= - \xi_{\mathrm{n}} v^{\rm (n)} _{ij} \Theta(\Delta_{ij}) 
	\end{align}
	with the normal velocity $v_{ij}^{\rm (n)} = (\frac{d}{dt} \bm{r}_i -\frac{d}{dt}  \bm{r}_j)\cdot \bm{n}_{ij}$ and the viscous constant $\xi_{\mathrm{n}}$, the diameter $d$, and the overlap length $\Delta_{ij} = d - r_{ij}$.
	The contact between particles is formed at $\Delta_{ij}=0$ ($r_{ij} = d$).
	Here, $\Theta(x)$ is the Heaviside step function satisfying $\Theta(x) = 1$ for $x \geq 0$ and $\Theta(x) = 0$ otherwise.
	The static force $F_{ij}^{(\mathrm{s})}$
	consists of the elastic contact force $F_{ij}^{(\mathrm{e})}$ and the attractive capillary force $F_{ij}^{(\mathrm{cap})}$ as 
	\begin{align}
		F_{ij}^{(\mathrm{s})} & = F_{ij}^{(\mathrm{e})} + F_{ij}^{(\mathrm{cap})}.
		\label{Fs}
	\end{align}
	The elastic contact force is given by
	\begin{align}
		F_{ij}^{(\mathrm{e})} &= k_{\mathrm{n}} \Delta_{ij} \Theta(\Delta_{ij}) \label{Fnel} 
	\end{align}
	with the spring constant $k_{\mathrm{n}}$.
	\begin{figure}[t]
		\centering
		\includegraphics[width=0.8\linewidth]{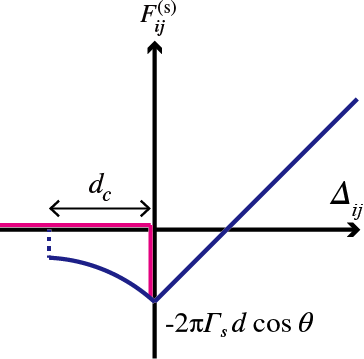}
		\caption{Static force $F_{ij}^{(\mathrm{s})}$ against $\Delta_{ij}$.
			The red line represents the behavior when the particles are approaching. 
			The blue line represents the behavior when the particles are separating. }
		\label{fig:wet_irr_force}
	\end{figure}
	The capillary force $F_{ij}^{(\mathrm{cap})}$ is modeled as \cite{roy2016micro,roy2017general,roy2018liquid,shi2020steady}
	\begin{align}\label{force_wet}
		F_{ij}^{(\mathrm{cap})} =
		\begin{cases}
			-2 \pi \Gamma_s d \cos \theta & \mbox{if } \Delta_{ij} \geq 0,\\[4pt]
			f^{\rm (a)}(\Hat{s}_{ij})\Theta\left (\Delta_{ij} +d_{\mathrm{c}} \right )  & \mbox{if } \Delta_{ij} < 0, \mbox{separation}, \\[4pt]
			0 & \mbox{if } \Delta_{ij} <0, \mbox{approaching}\\[4pt]
		\end{cases}
	\end{align}
	with
	\begin{align}
		f^{\rm (a)}(\Hat{s}_{ij}) = 
		\dfrac{-2 \pi \Gamma_s d \cos \theta}{1 + 1.05 \Hat{s}_{ij} + 2.5 \Hat{s}_{ij}^2}.
		\label{Fa}
	\end{align}
	Here, $\theta$ is the contact angle, $d_c$ is the rapture length of the capillary bridge, $\Hat{s}_{ij} = s_{ij}\sqrt{d/V_b}$ is the normalized separation distance, with the separation distance $s_{ij}=-\Delta_{ij}$. 
	The surface tension is denoted by $\Gamma_s$, which characterizes the strength of cohesion.
	The volume of the liquid bridge is denoted by $V_b$, which characterizes the rapture length $d_c$ as $d_c=(1+\theta/2)V_b^{1/3}$ \cite{lian1993theoretical,willett2000capillary}.
	When the particles approach before the contact, $F_{ij}^{(\mathrm{cap})}$ is zero. 
	After the particles contact at $\Delta_{ij} = 0$, the capillary bridge is formed, i.e., the attractive force becomes active.
	When the particles are separating, the force does not follow the same path; the attractive force is active until the capillary bridge disappears at $\Delta_{ij} = -d_c$.

	The static force $F^{(\mathrm{s})}$ given by Eqs. \eqref{Fs}-\eqref{Fa} is irreversible due to the capillary force, as shown in Fig. \ref{fig:wet_irr_force}. 
	When the particles are approaching before contact,  $F^{(\mathrm{s})}$ is zero and follows the red line.
	After the capillary bridge is formed at $\Delta_{ij}=0$, the capillary force $F^{(\mathrm{cap})}$ is active, and $F^{(\mathrm{s})}$ follows the blue line until the bridge is broken at $\Delta_{ij} = -d_c$.
	As shown in  Fig. \ref{fig:wet_irr_force}, the static force becomes zero at $\Delta_{ij} =\delta_0$ ($r_{ij} = d - \delta_0$) with $\delta_0 = 2\pi \Gamma_s d \cos \theta / k_{\mathrm{n}} > 0$, where positive $F_{ij}^{(\mathrm{e})}$ and negative $F_{ij}^{(\mathrm{cap})}$ cancel each other out.

	The particles are randomly placed with an initial packing fraction $\phi_{\mathrm{ini}}=0.45$ without any overlap.
	The system is gradually compressed until the packing fraction reaches a given value $\phi$.
	In each compression step, we increase the packing fraction by $\Delta \phi=0.000025$ with the affine transformation of the particle configuration and the system size.
	The particles are relaxed to a mechanical equilibrium state with the kinetic temperature $T<T_{\mathrm{th}}$ following Eqs. \eqref{eq:eom1} and \eqref{eq:eom2} with $\dot \gamma(t) = 0$.
	Here, the kinetic temperature is given by $T = \sum m | \bm{v}_i|^2 / (2N)$.
	
	After the compression, we apply the oscillatory shear strain as
	\begin{align}
		\gamma(t) = \gamma_0 \sin\omega t
	\end{align}
	for $N_{\mathrm{cyc}}$ cycles.
	Here, $\gamma_0$ and $\omega$ are the strain amplitude and the angular frequency, respectively, which are set small enough.
	In the last cycle, we measure the shear (storage) modulus $G$ as \cite{doi1988theory}
	\begin{align}
		G(\phi)=\frac{\omega}{\pi} \int_{0}^{2 \pi / \omega} \dd t \  \sigma_{xy}(\phi,t) \sin \omega t / \gamma_{0}
	\end{align}
	with the shear stress for a given $\phi$
	\begin{align}
		\sigma_{xy}(\phi,t) = - \dfrac{1}{2L^3} \sum_{i} \sum_{i < j}  (r_{ij,x} F_{ij,y} + r_{ij,y} F_{ij,x}). \label{S}
	\end{align}
	We also measure the pressure as
	\begin{align}
		P(\phi) = \dfrac{1}{3L^3} \sum_{i} \sum_{i < j}  \bm{r}_{ij} \cdot \bm{F}_{ij} \label{P}
	\end{align}
	after the last cycle,
	and calculate the bulk modulus as
	\begin{align}
		B(\phi) = \phi \frac{\dd P}{\dd \phi}.
	\end{align}
	Here, we have ignored the kinetic parts of the shear stress and the pressure because the contact stress is dominant in our dense system \cite{da2005rheophysics}.
	
	We use $N=3000$, $\gamma_0 = 1.0\times10^{-5}$, $\omega = 1.0\times10^{-4}
	\sqrt{k_{\mathrm n}/m}$, $N_\mathrm{cyc}=100$, and $T_{\mathrm{th}}=10^{-8}k_{\mathrm{n}}d_0^2$.
	We choose $d_c=5.0\times10^{-4}d_0$, $V_b=7.5\times10^{-11}d_0^3$, $\theta=\pi/9$, and
	$\Gamma_s/k_{\mathrm{n}} = 0, ~3.0\times10^{-3},~1.5\times10^{-2} ,~3.0\times10^{-2}$
	for the parameters of the capillary force
	following ref. \cite{roy2017general,roy2018liquid,shi2020steady}.
	Here, $\Gamma_s=0$ corresponds to dry particles.
	We adopt the Adams-Morton and Adams-Bashforth methods with a time step $\Delta t = 0.005\sqrt{m/k_{\mathrm{n}}}$ for the time evolution of $\bm{r}_i$ and $\bm{p}_i$, respectively.
	We have numerically confirmed that $N$ and $N_\mathrm{cyc}$ are large enough, and $\gamma_0$, $\omega$, $T_{\mathrm{th}}$, and $\Delta t$ are small enough not to influence our results. 
	
	
	\section{Mechanical properties}\label{sec:mechanical}
	
\begin{figure}[h]
		\centering
		\includegraphics[width=1.0\linewidth]{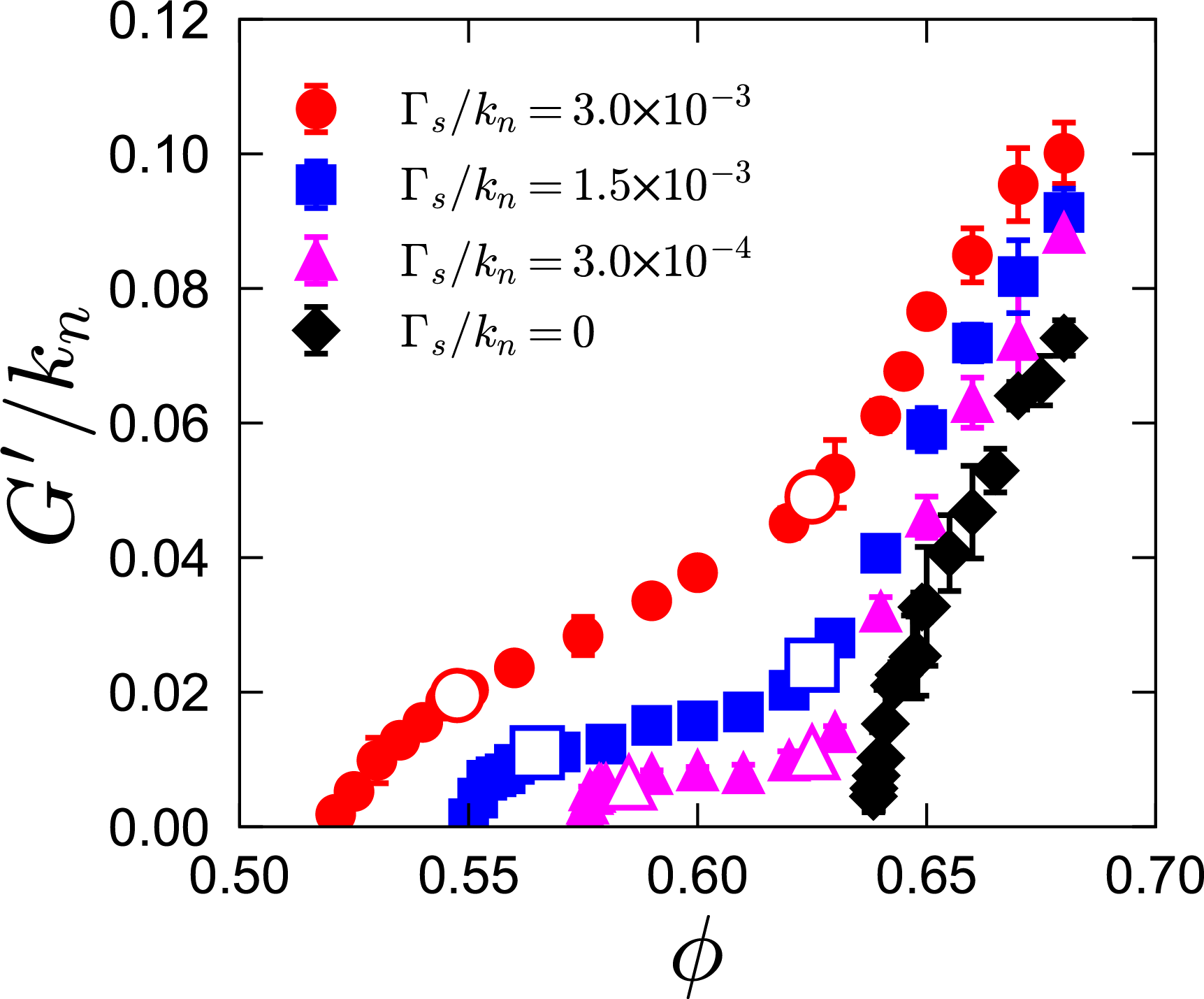}
		\caption{Shear modulus $G$ against $\phi$ for various $\Gamma_s$ with $\phi>\phi_c$. Open symbols represent inflection points.}
		\label{fig:phi-g}
	\end{figure}

	In Fig. \ref{fig:phi-g}, we plot the shear modulus $G$ against the volume fraction $\phi$ for various $\Gamma_s$. 
	For each $\Gamma_s$, $G$ becomes non-zero as the packing fraction $\phi$ exceeds a critical fraction $\phi_c$. 
	As $\Gamma_s$ increases, the critical fraction $\phi_c$ decreases, as shown in \ref{sec:phic}.
	The shear modulus $G$ increases with $\phi$.
	For wet particles with $\Gamma_s>0$, there are two inflection points (open symbols) where the curvature of $G(\phi)$ changes sign with $\frac{\dd[2]}{\dd\phi^2}G(\phi) = 0$.
	The position of the inflection point with higher $\phi$ is almost independent of $\Gamma_s$ ($\phi \simeq 0.63$).
	The inflection point does not exist for dry particles with $\Gamma_s = 0$.
	The inflection points indicate that the simple power law scaling for repulsive particles \cite{ohern2002random,ohern2003jamming} is not satisfied for wet granular materials.
	Similar behaviors are reported for two-dimensional particles with a simple reversible attractive interaction \cite{Koeze2020Elasticity}, but only one inflection point exists in the system.
	
	\begin{figure}[h]
		\centering
		\includegraphics[width=1.0\linewidth]{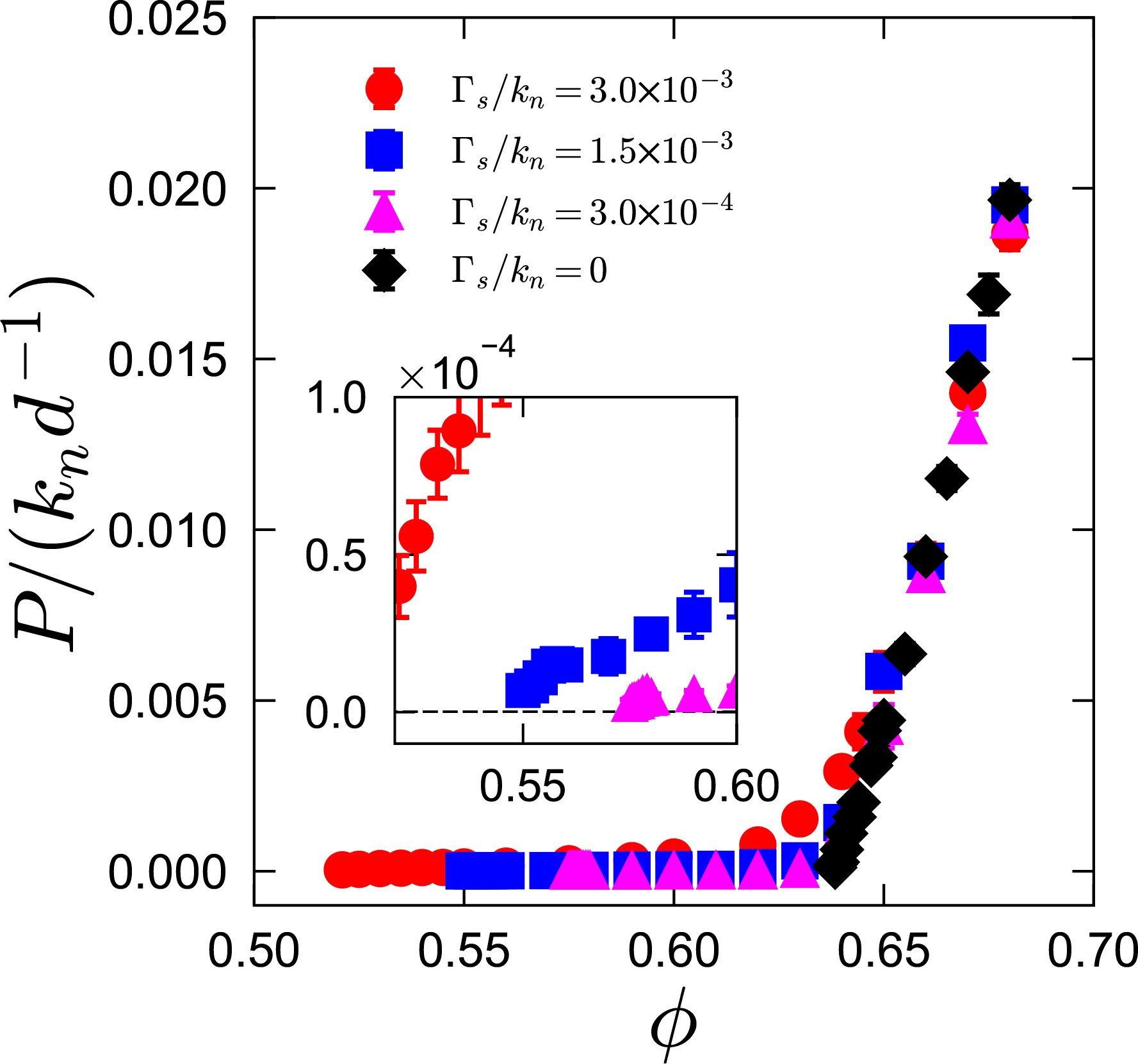}
		\caption{Pressure $P$ against $\phi$ for various $\Gamma_s$ with $\phi>\phi_c$. The inset shows $P$ in the vicinity of $\phi_c$. The dashed line represents $P=0$.}
		\label{fig:phi-P}
	\end{figure}
	
	Figure \ref{fig:phi-P} displays the pressure $P$ against the volume fraction $\phi$ for various $\Gamma_s$ with $\phi>\phi_c$.
	The inset of Fig. \ref{fig:phi-P} shows $P$ near $\phi_c$.
	For each $\Gamma_s$, $P$ increases with $\phi$.
	The pressure is positive and almost $0$ even in the vicinity of $\phi_c$, as shown in the inset of Fig. \ref{fig:phi-P}. 
	For high $\phi>0.65$, $P$ seems independent of $\Gamma_s$.
	See \ref{sec:critical} for the relation between $G$ and $P$.
	
	\begin{figure}
		\centering
		\includegraphics[width=1.0\linewidth]{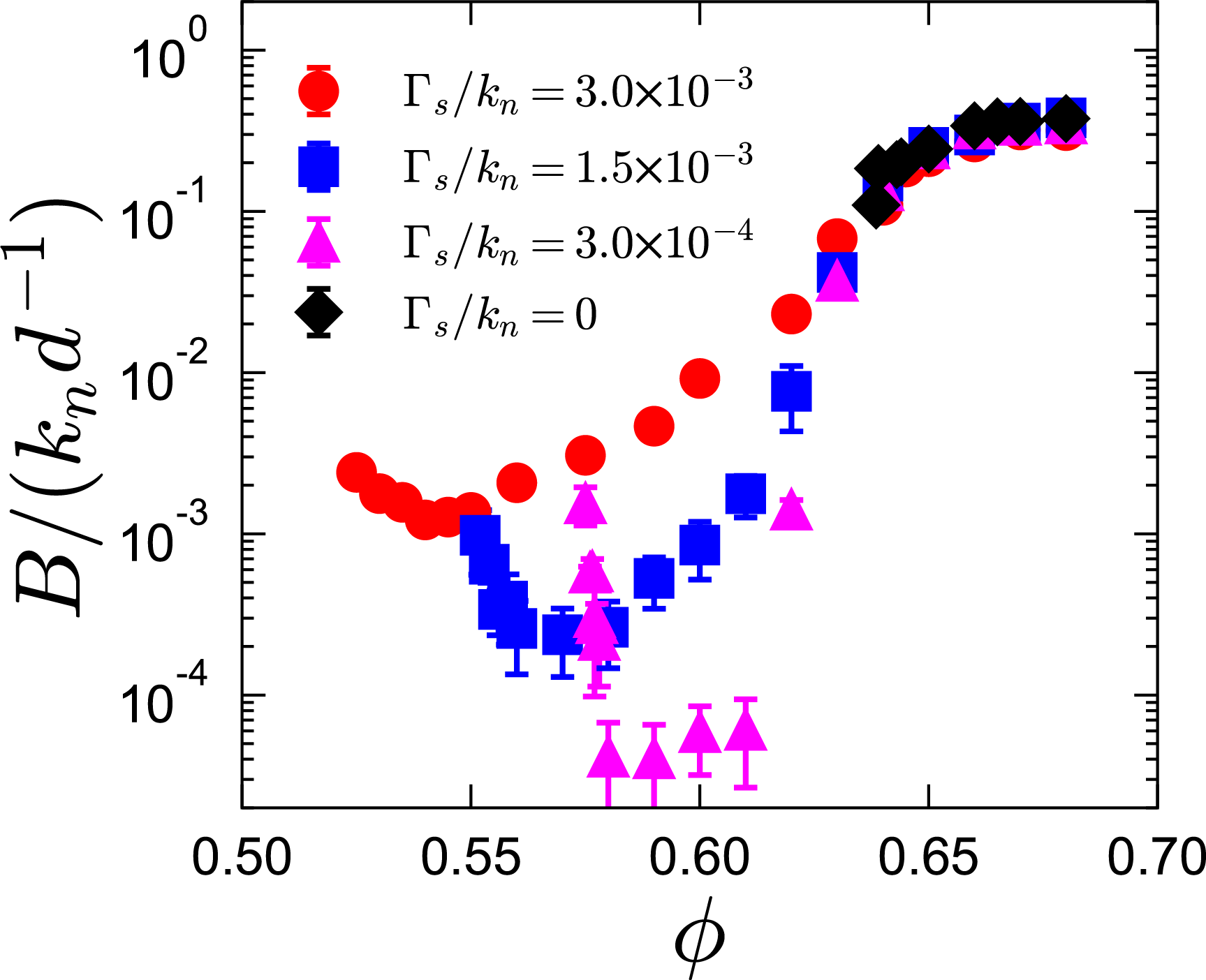}
		\caption{Bulk modulus $B$ against $\phi$ for various $\Gamma_s$ with $\phi>\phi_c$.}
		\label{fig:phi-B}
	\end{figure}
	
	In Fig. \ref{fig:phi-B}, we demonstrate the bulk modulus $B$ against the packing fraction $\phi$ for various $\Gamma_s$ with $\phi>\phi_c$.
	The bulk modulus $B$ is not a monotonic function of $\phi$ for $\Gamma_s>0$, while $B$ for $\Gamma_s=0$ does not exhibit such non-monotonic behavior.
	As the packing fraction $\phi$ decreases, $B$ rapidly increases near $\phi_c$, which is not shown in the previous study \cite{Koeze2020Elasticity}.
	The non-monotonic behavior in $B$ of wet granular materials indicates that $B$ does not obey the power law scaling for dry repulsive particles \cite{ohern2002random,ohern2003jamming}.
	
	\section{Geometrical properties}\label{sec:geometrical}
	
	In this section, we analyze the geometrical properties of wet granular materials.
	In Sect. \ref{sec:Z}, we show the $\phi$-dependence of the coordination number.
	Section \ref{sec:Pair} demonstrates the change in the pair correlation function for $\Gamma_s>0$.
	In Sect. \ref{sec:Voro}, we discuss the probability density function for the volume of the Voronoi cell obtained by the Voronoi tessellation.

	\subsection{Coordination number}
	\label{sec:Z}
\begin{figure}[h]
		\centering
		\includegraphics[width=1.0\linewidth]{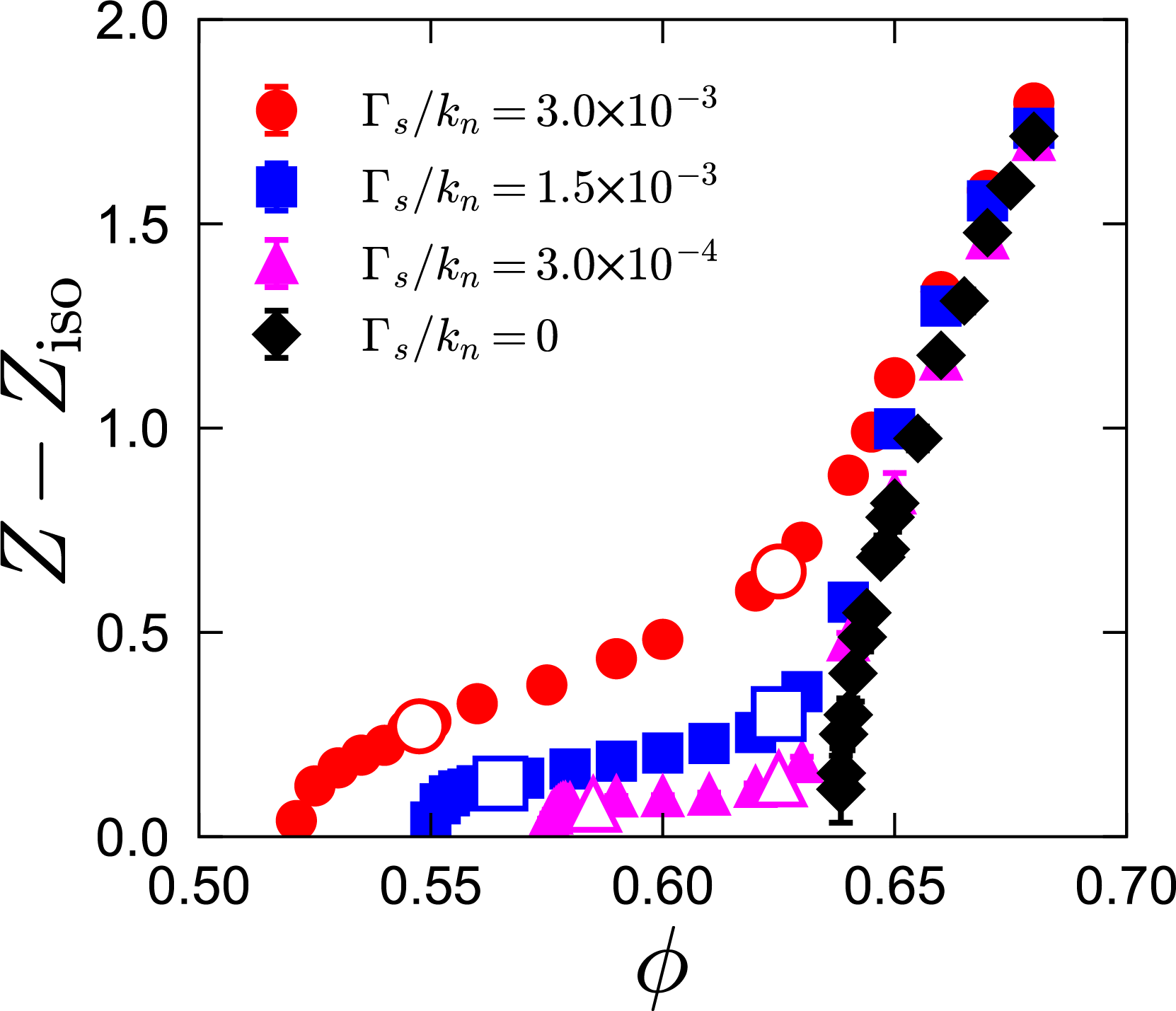}
		\caption{Excess coordination number $Z-Z_{\mathrm{iso}}$ against $\phi$ for various $\Gamma_s$ with $\phi>\phi_c$.
			Open symbols represent the inflection points.}
		\label{fig:phi-z-z_iso}
	\end{figure}

	We plot the excess coordination number $Z-Z_{\mathrm{iso}}$ against the volume fraction $\phi$ for various $\Gamma_s$ with $\phi>\phi_c$ in Fig. \ref{fig:phi-z-z_iso}.
	Here, we calculate the coordination number $Z$ after the final oscillatory shear as
	\begin{align}
		Z = 2 N_{\mathrm{con}}/N,
	\end{align}
	where $N_{\mathrm{con}}$ is the total number of contacts with $F_{ij} \neq 0$.
	For three-dimensional frictionless particles, the isostatic value $Z_{\mathrm{iso}}$ equals $6$ \cite{van2009jamming}.
	As $\phi$ increases, so does $Z-Z_{\mathrm{iso}}$, increasing from $0$.
	There are two inflection points (open symbols) in $Z-Z_{\mathrm{iso}}$ for each $\Gamma_s$. 
	Their positions are almost the same as those for $G$ in Fig. \ref{fig:phi-g}.
	The existence of two inflection points is natural if the relation $G \propto Z-Z_{\mathrm{iso}}$ for repulsive particles \cite{ohern2002random} holds for wet granular particles, which is discussed in \ref{sec:critical}.

	\subsection{Pair correlation function}
	\label{sec:Pair}
	\begin{figure*}[t]
		\centering
		\includegraphics[width=1.0\linewidth]{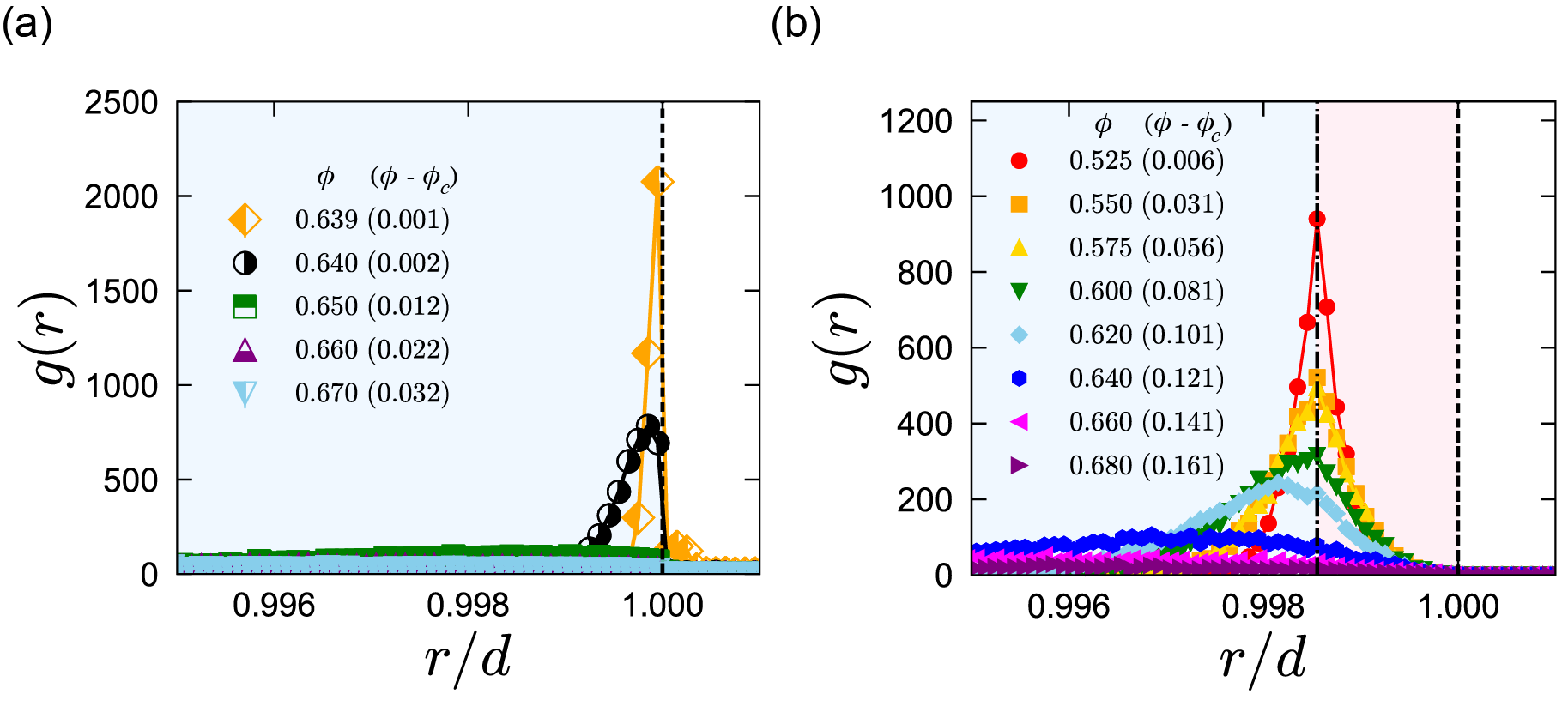}
		\caption{Pair correlation function $g(r)$ against $r$ for various $\phi$ with $\Gamma_s/k_{\mathrm{n}} = 0$ (a) and $\Gamma_s/k_{\mathrm{n}} = 3.0\times 10^{-3}$ (b). Dashed and dash-dotted lines represent $r = d$ and $r=d-\delta_0$, respectively. The blue shaded region corresponds to $r<d-\delta_{0}$. 
			The red shaded region corresponds to $d-\delta_{0}\leq r \leq d+d_c$.
		}
		\label{fig:g_r}
	\end{figure*}
	\begin{figure}[h]
		\centering
		\includegraphics[width=1.0\linewidth]{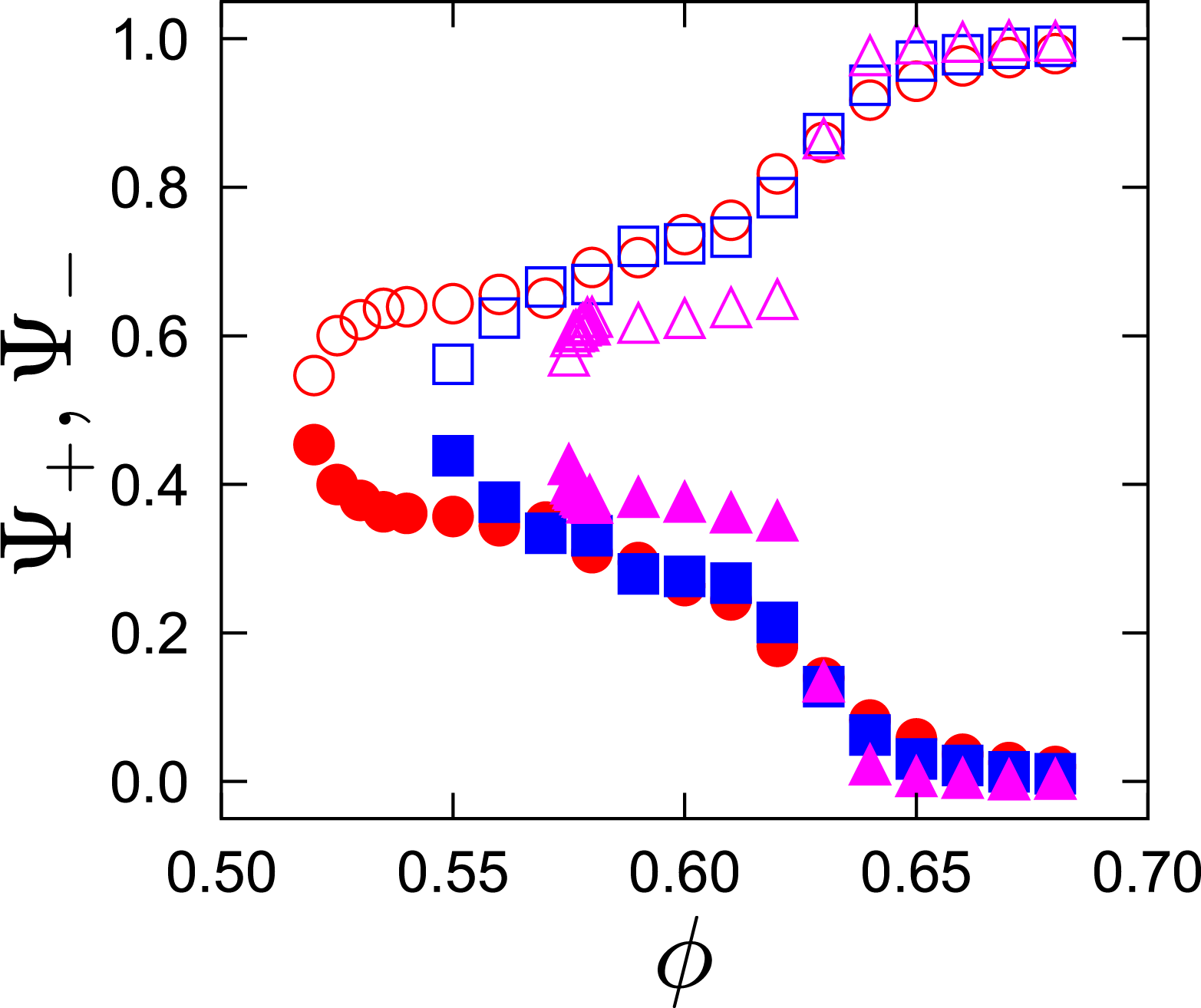}
		\caption{Plot of $\Psi_\mathrm{+}$ and $\Psi_\mathrm{-}$ against $\phi$ for various $\Gamma_s$ with $\phi>\phi_c$.
			Circles , squares, and triangles represent $\Gamma_s = 3.0\times 10^{-3}$, $\Gamma_s = 1.5\times 10^{-3}$, and $\Gamma_s = 3.0\times 10^{-4}$ respectively. 
			Open (filled) symbol corresponds to $\Psi_\mathrm{+}$ ($\Psi_\mathrm{-}$).
		}
		\label{fig:red-blue}
	\end{figure}
	
	In Fig. \ref{fig:g_r}, we demonstrate the pair correlation function $g(r)$ against $r$ in the static state after the final oscillatory shear for various $\phi>\phi_c$. 
	The pair correlation function $g(r)$ is given by
	\begin{align}
		g(r) = \dfrac{L^3}{N^2} \left\langle \sum_{i}\sum_{j \neq i} \delta^3(r -r_{ij}) \right\rangle.
		\label{pcf}
	\end{align}
	The blue area in Fig. \ref{fig:g_r} ($r< d-\delta_0$) corresponds to the region with the static force $F_{ij}^{(\mathrm{s})}>0$, while the red area ($d-\delta_0 < r < d+d_c$) represents the region with $F_{ij}^{(\mathrm{s})}<0$. 
	The first peak of $g(r)$ rapidly increases as $\phi \to \phi_c$.
	For $\Gamma_s/k_{\mathrm{n}} = 0$ (Fig. \ref{fig:g_r} (a)), the peak position approaches $r=d$ as $\phi\to \phi_c$ \cite{silbert2006structural}.
	For $\Gamma_s/k_{\mathrm{n}} = 3.0\times 10^{-3}$ (Fig. \ref{fig:g_r} (b)), the peak position decreases to $r=d-\delta_0$, at which $F_{ij}^{(\rm s)} = 0$. 
	We should note that the gel-like structure characterized by power law decay in the structure factor \cite{zheng2016shear} is not observed in our simulation with $\phi>\phi_c$.

	In Fig. \ref{fig:g_r} (b), $g(r)$ has a large value even in the red region with $d-\delta_0<r<d+d_c$ near $\phi_c$, which indicates that many contacts have a negative static force $F_{ij}^{(\mathrm{s})}$.
	Here, we introduce $N_+$ and $N_-$ as the number of contacts with $F_{ij}^{(\mathrm{s})}>0$ and $F_{ij}^{(\mathrm{s})}<0$, respectively, and plot the ratios
	$\Psi_{+}=N_{+}/N_{\mathrm{con}}$ and $\Psi_{-}=N_{-}/N_{\mathrm{con}}$ with the total number of contacts $N_{\mathrm{con}} = N_+ + N_-$ against $\phi$ for various $\Gamma_s>0$ in Fig. \ref{fig:red-blue}.
	Note that
	$N_+ = \int_{0}^{d - \delta_{0}}g(r) \dd r$ and $N_- = \int^{d+d_c}_{d-\delta_{0}}g(r) \dd r$ if all pairs with $r_{ij} < d-\delta_{0}$ are contacting.
	For high $\phi$, $F_{ij}^{(\mathrm{s})}$ is positive for almost all contacts ($\Psi_{+} = 1$ and $\Psi_{-} = 0$), which is consistent with high $P$ for large $\phi$ in Fig. \ref{fig:phi-P}.
	As $\phi$ decreases to $\phi_c$, $\Psi_{+}$ and $\Psi_{-}$ approach $0.5$. 
	This indicates that the pressure $P$ of wet granular materials decreases as $\phi \to \phi_c$ because the number of contacts with $F_{ij}^{(\mathrm{s})}<0$ increases.
	This behavior is different from that of dry particles with purely repulsive interaction, where the positive contact force decreases to $0$, keeping $N_+ = N_{\mathrm{con}} \simeq Z_{\rm iso} N / 2$ ($\Psi_+=1$).

	\subsection{Voronoi tessellation}
	\label{sec:Voro}
	
	The structure of disordered particles has been studied using the Voronoi tessellation \cite{bernal1964bakerian,finney1970random,finney1970random2,oger1996voronoi,jullien1996computer,yang2002voronoi,aste2007invariant,xu2007analysis}.
	A Voronoi cell associated with each particle contains an ensemble of points closer to a given sphere center than any other.
	In ref. \cite{xu2007analysis}, it is reported that the probability density function for the volume of the Voronoi cell changes due to the capillary force, but the $\phi$-dependence is not investigated.
	
	\begin{figure}[h]
		\centering
		\includegraphics[width=1.0\linewidth]{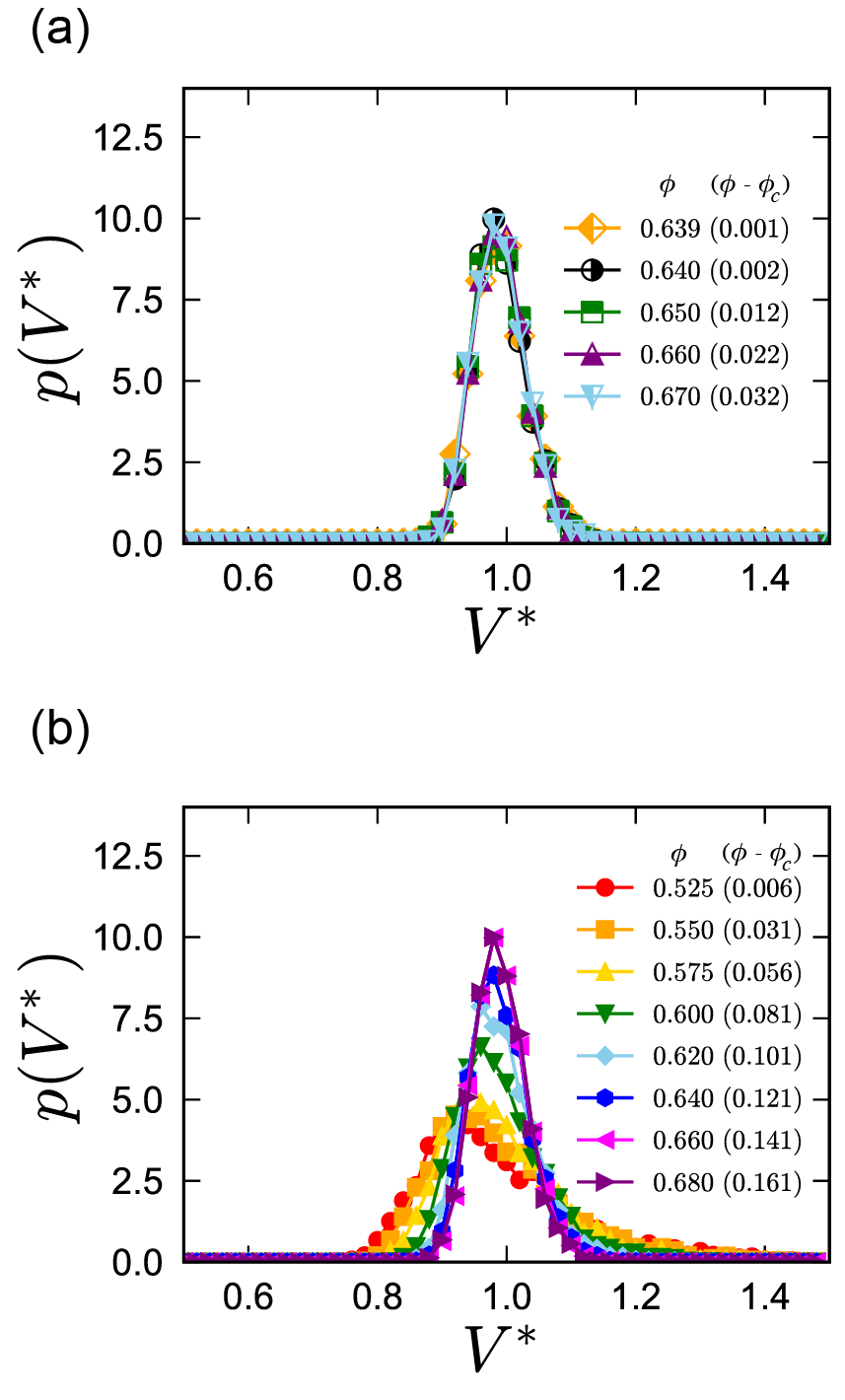}
		\caption{Probability density function $p(V^*)$ against $V^*$ with $\Gamma_s/k_{\mathrm{n}}  = 0$ (a) and $\Gamma_s/k_{\mathrm{n}}  = 3.0\times 10^{-3}$ (b) for various $\phi>\phi_c$. 
		}
		\label{fig:pdf_voro}
	\end{figure}
	We obtain the Voronoi cell from the particle configuration after the final oscillatory shear using the VORO++ code library \cite{rycroft2009voro++}.
	We define the volume of the Voronoi cell as $V$ and plot the probability density functions $p(V^{*})$ of the normalized volume $V^{*} = V/\Bar{V}$ with the average of the volume $\Bar{V} = L^3/N$ in Fig. \ref{fig:pdf_voro} for different $\phi>\phi_c$.
	Note that $\Bar{V}$ is related to the packing fraction $\phi$ as $\Bar{V} = \pi d^3 /(6\phi)$.
	For dry particles ($\Gamma_s/k_{\mathrm{n}}  = 0$),
	the probability density function is almost independent of $\phi$ (Fig. \ref{fig:pdf_voro}(a)).
	This independence might indicate that the volume of the Voronoi cell is affinely deformed by changing $\phi$.
	For wet particles ($\Gamma_s/k_{\mathrm{n}}  = 3.0\times 10^{-3}$), the width of the probability density functions increases, and the peak position of $p(V^{*})$ is shifted to lower $V^{*}$ as $\phi$ decreases (Fig. \ref{fig:pdf_voro}(b)), which is consistent with an experiment of wet particles \cite{xu2007analysis}.

	\begin{figure}[h]
		\centering
		\includegraphics[width=1.0\linewidth]{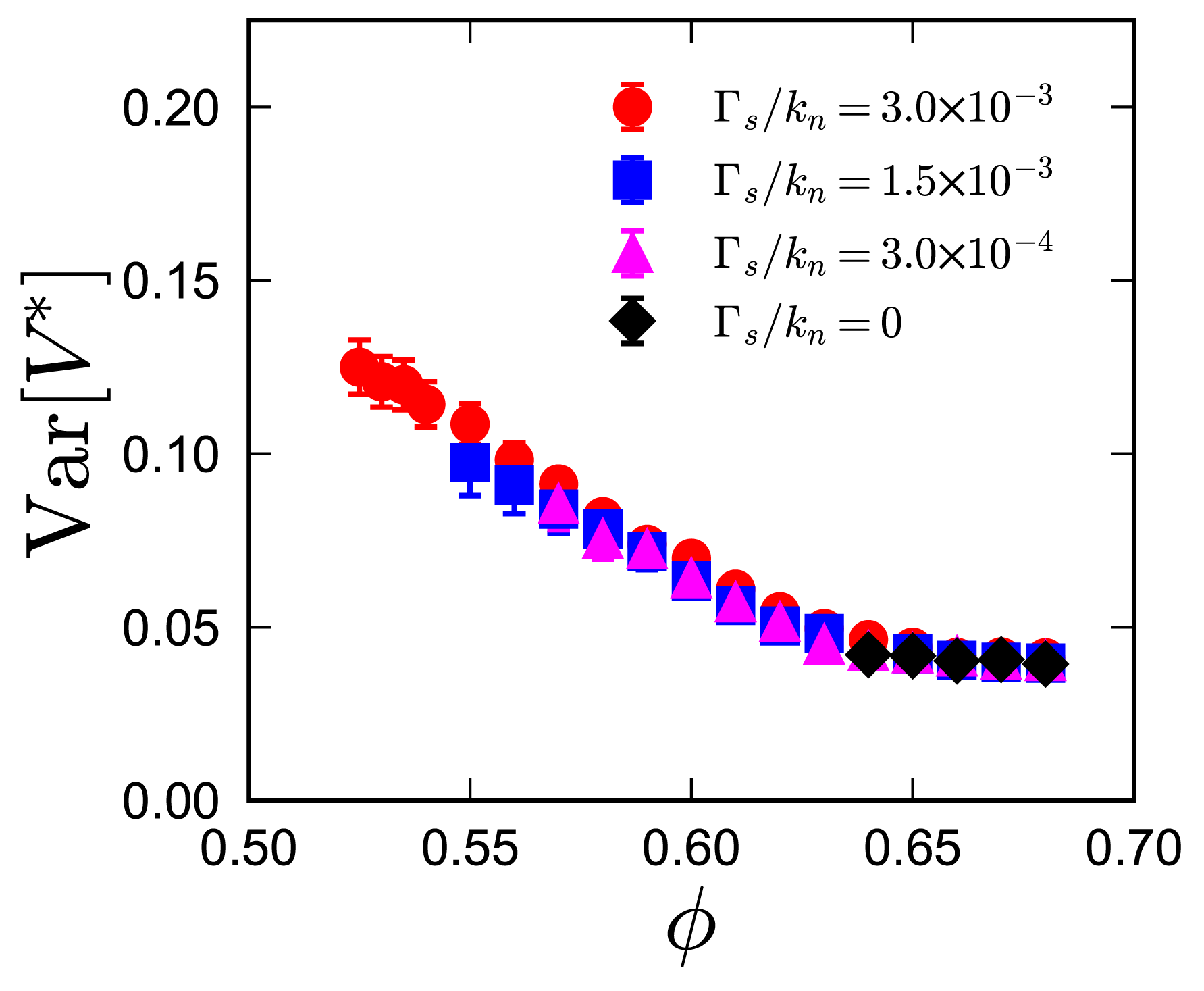}
		\caption{Variance of $V^*$ against $\phi$ for various $\Gamma_s$ with $\phi>\phi_c$.
		}
		\label{fig:voro_sigma2}
	\end{figure}

	In Fig. \ref{fig:voro_sigma2}, the variance of the normalized volume $V^{*}$ is plotted against $\phi$ for different $\Gamma_s$ with $\phi>\phi_c$.
	The variances for different $\Gamma_s$ collapse onto a master curve.
	However, the range of $\phi$ for each $\Gamma_s$ depends on $\phi_c$ and broadens as $\Gamma_s$ increases.
	For $\Gamma_s/k_{\mathrm n}=0$,
	the variance is almost independent of $\phi$,
	which is consistent with the probability density function of $V^*$ in Fig. \ref{fig:pdf_voro} (a).
	However, the variance for $\Gamma_s >0$ increases as $\phi$ approaches $\phi_c$.
	This corresponds to the increase in the width of $p(V^{*})$ in Fig. \ref{fig:pdf_voro} (b).
	The packing fraction where the variance increases is close to the inflection point for $G$ ($\phi \simeq 0.63$) in Fig. \ref{fig:phi-g}.

	\section{Conclusion and discussion}\label{sec:Conclusion}
	We numerically studied the mechanical and geometrical properties of wet granular materials with $\phi>\phi_c$.
	For $\Gamma_s>0$, the shear modulus $G(\phi)$ has two inflection points, and the bulk modulus $B(\phi)$ exhibits a non-monotonic behavior.
	These mechanical properties are qualitatively different from those for dry particles with $\Gamma_s = 0$, where $G(\phi)$ and $B(\phi)$ obey simple power law scalings near $\phi_c$
	\cite{ohern2002random,ohern2003jamming}.
	The excess coordination number $Z(\phi)-Z_{\mathrm{iso}}$ also has two inflection points.
	The peak position in the pair correlation function $g(r)$ becomes lower than the diameter $d$ due to the attractive capillary force.
	The probability density function for the volume of the Voronoi cell broadens as the packing fraction approaches $\phi_c$.
	These results indicate that the geometrical properties change with the mechanical properties due to the capillary force.

	The breakdown of simple power-law behaviors in $G(\phi)$ and $B(\phi)$ has been reported in ref. \cite{Koeze2020Elasticity} for two-dimensional particles with a simple reversible attractive force.
	However, it is also reported that $G$ and $B$ satisfy critical scaling laws, including the strength of attraction.
	The attractive force in ref. \cite{Koeze2020Elasticity} differs from the irreversible capillary force, and the critical scaling laws do not apply to the mechanical properties shown in this paper because of the two inflection points for $G(\phi)$ and the non-monotonic behavior in $B(\phi)$, which are not observed in ref. \cite{Koeze2020Elasticity}.
	The critical scaling laws near $\phi_c$ in wet granular materials will be the subject of future study.

	In this study, we have neglected contact friction between particles to focus on the effect of the attractive capillary force.
	Recent studies have reported that 
	the contact friction affects the critical behaviors near $\phi_c$
	for dry repulsive particles  \cite{somfai2007critical,silbert2010jamming,otsuki2017discontinuous,otsuki2021shear}.
	However, it is unclear whether the friction force changes the mechanical properties of attractive wet particles shown in this study.
	Further work is necessary to resolve this issue.
	
	\section*{Acknowledgment}
	K. Y. and M.O. thank S. Takada, T. Nakamura, and H. Mizuno for helpful discussions. 
	Numerical computation in this work was conducted at the Yukawa Institute Computer Facility. 
	We would like to thank Editage (www.editage.com) for English language editing.
	K.Y. is partially supported by Leave a Nest Co., Ltd., Hosokawa Powder Technology Foundation (Grant No.\ HPTF20506), and the Grant-in-Aid for Japan Society for Promotion of Science JSPS Research Fellow (Grant No.\ 21J13720).
	M.O. is partially supported by Scientific Grant-in-Aid of Japan Society for the Promotion of Science, KAKENHI (Grants No.\ 19K03670 and No.\ 21H01006).
	
	\section*{Author contribution statement}
	K.Y. carried out the numerical simulations. K.Y. and M.O. interpreted the results and wrote the manuscript.
	\appendix
	\section{Critical fraction}\label{sec:phic}
	
\begin{figure}[h]
		\centering
		\includegraphics[width=1.0\linewidth]{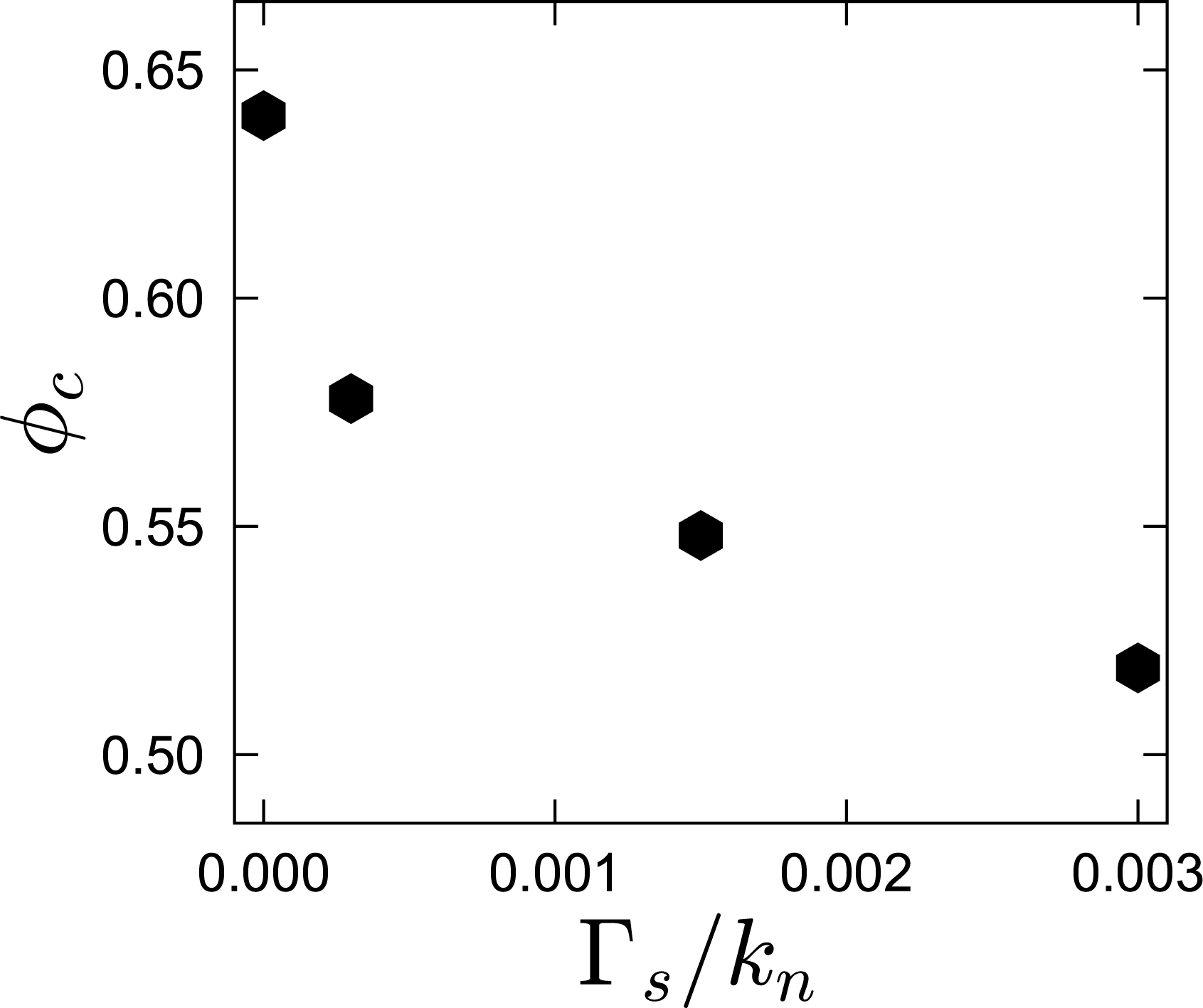}
		\caption{Critical fraction $\phi_c$ against $\Gamma_s$.
		}
		\label{fig:phic}
	\end{figure}

	This appendix shows the $\Gamma_s$-dependence of the critical fraction $\phi_c$.
	Here, we define the critical fraction $\phi_c$ as the packing fraction where $G$ exceeds a threshold $G_{\mathrm{th}}$ with $G_{\mathrm{ th}}/k_{\mathrm n}=1.0\times10^{-4}$.
	We have checked that $\phi_c$ does not change if we use a smaller $G_{\mathrm{th}}/k_{\mathrm n}=5.0\times10^{-5}$.
	In Fig. \ref{fig:phic}, we plot the critical fraction $\phi_c$ against $\Gamma_s$.
	The critical fraction $\phi_c$ decreases as $\Gamma_s$ increases, which is consistent with the results of ref. \cite{Koeze2020Elasticity}.
	
	\section{Critical scaling of $G$}\label{sec:critical}
	
	\begin{figure}[h]
		\centering
		\includegraphics[width=1.0\linewidth]{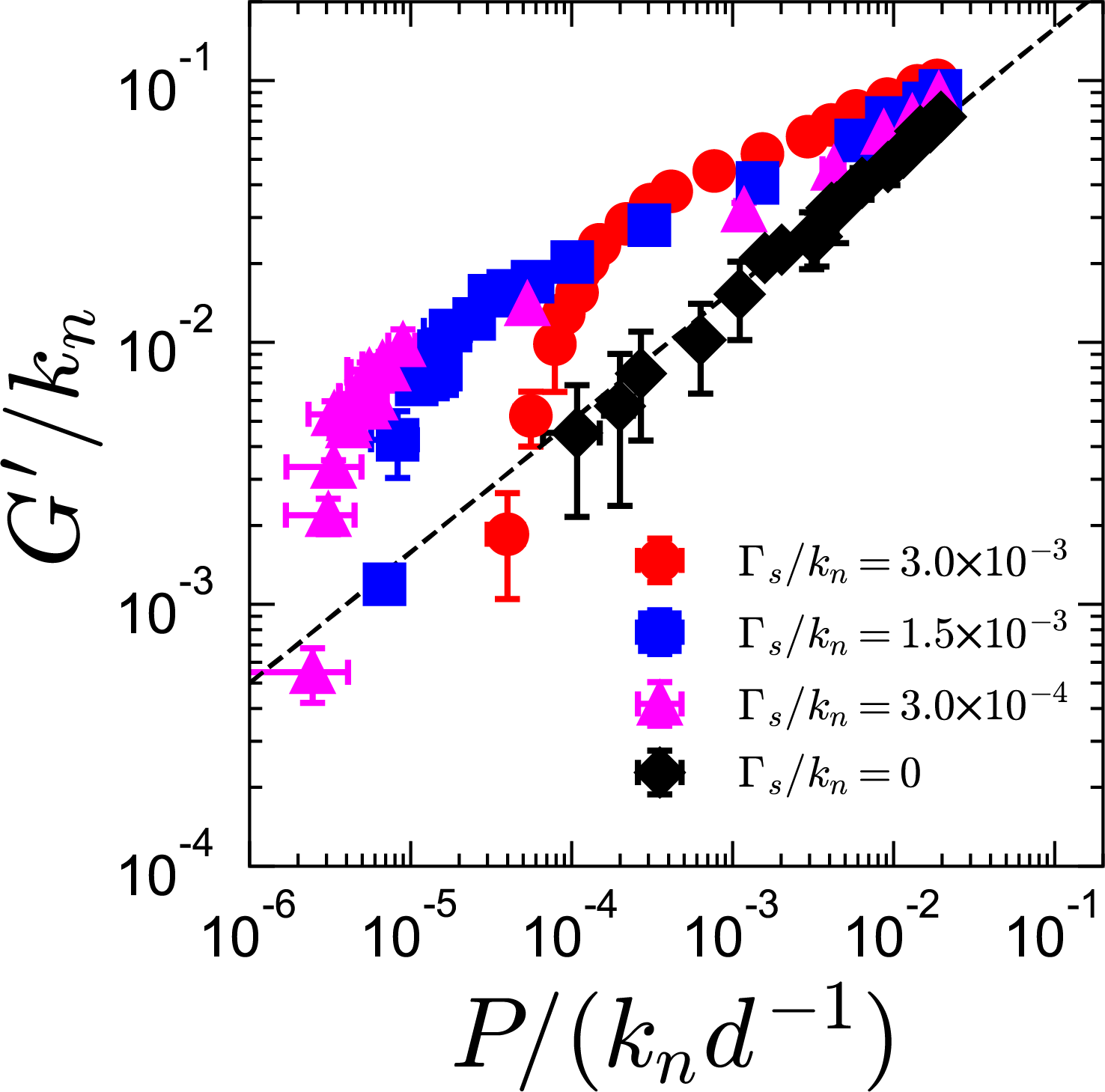}
		\caption{Shear modulus $G$ against $P$ for various $\Gamma_s$. The dashed line represents $G \propto P^{1/2}$.}
		\label{fig:P-G}
	\end{figure}
	
	In this appendix, we numerically investigate the effect of the capillary force on the scaling laws for $G$ of dry repulsive particles.
	For frictionless particles with the linear elastic interaction $F_{ij}^{\mathrm{(e)}}$ given by Eq. \eqref{Fnel},
	$G$ satisfies $G \propto P^\alpha$ with $\alpha \simeq 1/2$ and $G \propto  (Z-Z_{\mathrm{iso}})$
	\cite{ohern2002random,ohern2003jamming}.

	Figure \ref{fig:P-G} displays $G$ against $P$ for various $\Gamma_s$ obtained from the data in Figs. \ref{fig:phi-g} and \ref{fig:phi-P}.
	For dry particles with $\Gamma_s = 0$, $G$ almost satisfies the scaling law $G \propto P^\alpha$ with $\alpha = 1/2$. 
	For wet particles with $\Gamma_s>0$, there is a region where $G$ behaves as a power law function with an exponent lower than $1/2$ for $P/(k_{\mathrm{n}} d^{-1}) > 10^4$. 
	However, the power-law behavior seems to break down as $P$ decreases, and $G$ seems to become zero at a finite $P$.

\begin{figure}[htb]
		\centering
		\includegraphics[width=1.0\linewidth]{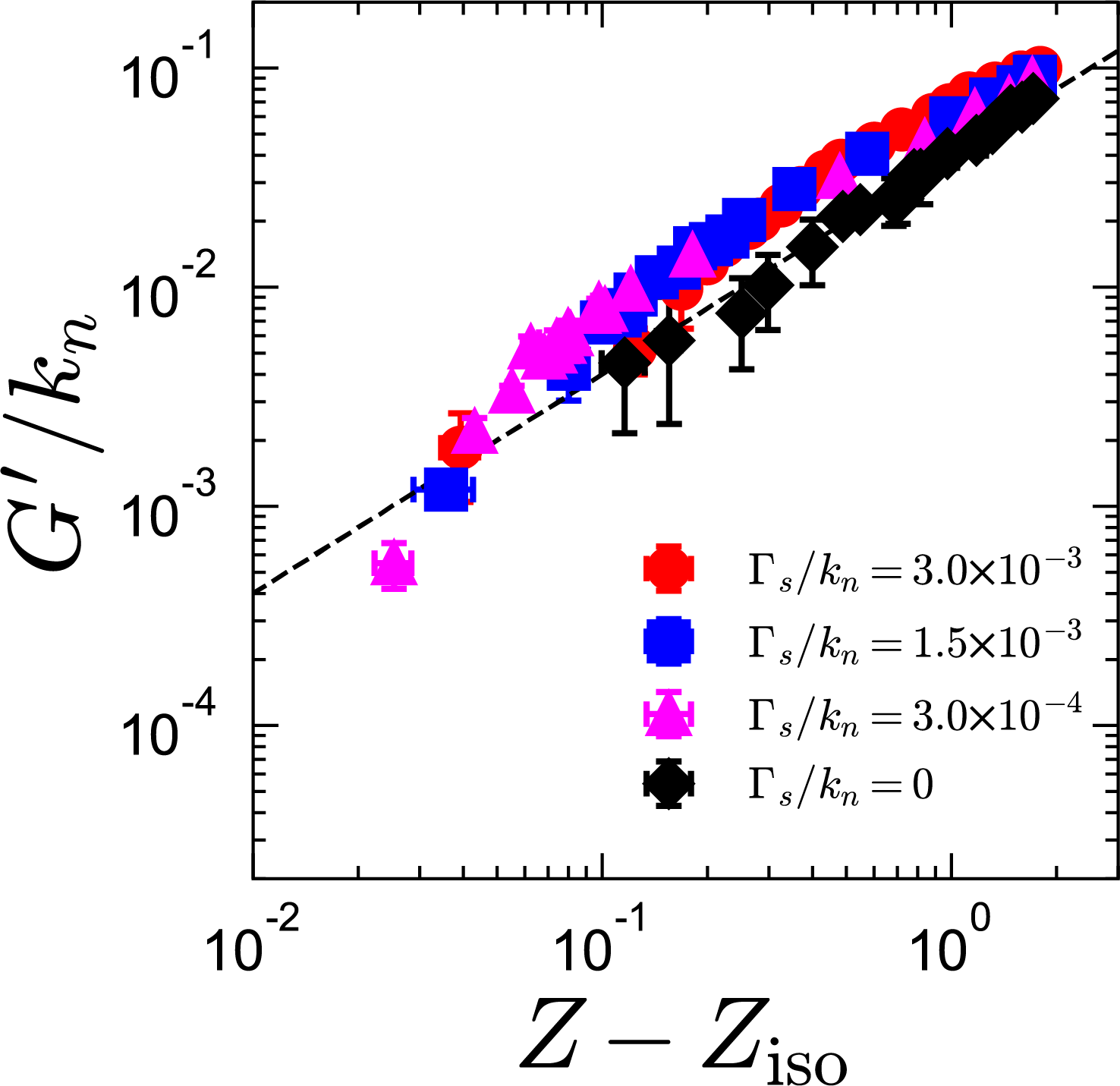}
		\caption{Shear modulus $G$ against $Z-Z_{\mathrm{iso}}$ for various $\Gamma_s$. The dashed line represents $G \propto Z-Z_{\mathrm{iso}}$.}
		\label{fig:z-z_iso-g}
	\end{figure}

	In Fig. \ref{fig:z-z_iso-g}, we show
	the shear modulus $G$ against the excess coordination number $Z-Z_{\mathrm{iso}}$ for different $\Gamma_s$ obtained from the data in Figs. \ref{fig:phi-g} and \ref{fig:phi-z-z_iso}.
	For dry particles with $\Gamma_s=0$, $G \propto  (Z-Z_{\mathrm{iso}})$ is satisfied.
	For $\Gamma_s > 0$, there is a region where $G$ is proportional to $Z-Z_{\mathrm{iso}}$, but the scaling relation is broken for smaller $Z-Z_{\mathrm{iso}}$ near $\phi_c$.
	
	\bibliographystyle{spphys}

\end{document}